\documentclass{article}%
\usepackage{amsmath}%
\usepackage{amsfonts}%
\usepackage{amssymb}%
\usepackage{graphicx}

\usepackage{subfig}
\usepackage{sidecap}
\usepackage{authblk}
\usepackage{epstopdf}
\usepackage{hyperref}
\usepackage{xcolor}
\usepackage{float}
\usepackage{cite}
\usepackage{url}
\definecolor{dark-red}{rgb}{0.,0.,0}
\definecolor{dark-blue}{rgb}{0.,0.,1}
\definecolor{medium-blue}{rgb}{0,0,1}
\hypersetup{
    colorlinks, linkcolor={dark-red},
    citecolor={dark-blue}, urlcolor={medium-blue}
}

\begin{document}

\title{Effects of the tensor force on the ground and first $2^{+}$ states of the magic $^{54}$Ca nucleus}
\author[1,2]{E. Y{\"{u}}ksel}
\author[2]{N. Van Giai}
\author[2]{E. Khan}
\author[1,2]{K. Bozkurt}

\affil[1]{Physics Department,Yildiz Technical University, 34210
Esenler, Istanbul, Turkey}
\affil[2]{Institut de Physique Nucl\'eaire, Universit\'e Paris-Sud,
IN2P3-CNRS, F-91406 Orsay Cedex, France}

\date{\today}
\maketitle
\begin{abstract}
The magic nature of the $^{54}$Ca nucleus is investigated in the light of the recent experimental results. We employ both HFB and HF+BCS methods using Skyrme-type SLy5, SLy5+T and T44 interactions. The evolution of the single-particle spectra is studied for the N=34 isotones: $^{60}$Fe, $^{58}$Cr, $^{56}$Ti and $^{54}$Ca. An increase is obtained in the neutron spin-orbit splittings of $p$ and $f$ states due to the effect of the tensor force which also makes $^{54}$Ca a magic nucleus candidate. QRPA calculations 
on top of HF+BCS are performed to investigate the first $J^{\pi}$=$2^{+}$ states of the calcium isotopic chain. A good agreement for excitation energies is obtained when we include the tensor force in the mean-field part of the calculations. The first $2^{+}$ states indicate a subshell closure for both $^{52}$Ca and $^{54}$Ca nuclei. We confirm that the tensor part of the interaction is quite essential in explaining the neutron subshell closure in $^{52}$Ca and $^{54}$Ca nuclei.
\end{abstract}

\section{Introduction}
One of the most important steps in the study of the atomic nuclei was the discovery of the nuclear magic numbers \cite{mgm49,hjs49}.
 Despite the great success of the spin-orbit force in explaining the closed shells and subshells in the nuclear stability valley, there is a long standing controversy about its adequacy near the drip lines. In recent years, both theoretical and experimental results indicate the formation of new magic numbers \cite{lid04,frid05,ga06,sor08,kan09,hof09,step13}
and it seems that a large part of our knowledge about the location of magic numbers should be renewed in the near future.

Over the last few decades, there is an ongoing argument about the nature of the $^{54}$Ca nucleus \cite{hagen12,rej07,cor09,hol13,hol12}. However, these studies could not reach a common conclusion about the magicity of the $^{54}$Ca nucleus. One of the appropriate theoretical approach to predict nuclear shell effects is the self-consistent mean field method. This is the purpose of the present work.

In a recent experiment, very striking results were obtained by D. Steppenbeck et al. \cite{step13}. The data from the spectroscopic study of the neutron-rich $^{54}$Ca nucleus using proton knock-out reactions reveal the doubly magic nature of this nucleus. The first $2^{+}$ state is found at about 2 MeV, thus indicating a magic structure where the neutron $2p_{3/2}-2p_{1/2}$ subshells would be occupied whereas the neutron $1f_{5/2}$ level would be empty. In order to get a clear understanding of the experimental results, theoretical predictions are thus needed. 

In the last decade, it has been shown that the tensor force plays an active role in the shell evolution of nuclei. The tensor component of the effective N-N interaction contributes also to the spin-orbit part of the nuclear mean field and thus, it can affect considerably the single-particle (s.p.) energies \cite{ot05,colo07}. In the Hartree-Fock (HF) type of description, the tensor force is either added directly to the known Skyrme interactions (e.g., SLy5+T \cite{colo07}) or all Skyrme terms including the tensor component are refitted on the empirical ground state properties (e.g., the TIJ family of force parameters, see Ref. \cite{les07}). Detailed explanations and more information about the tensor force can be found in Refs. \cite{ot05,colo07}. Recently, adding the tensor term to the known
SLy5 force could explain well the experimental energy differences between the s.p. states of Z=50 isotopes and N=82 isotones \cite{colo07}. In addition, the emergence of the magicity in the $^{52}$Ca and $^{54}$Ca nuclei occurs when Skyrme-Hartree-Fock calculations are performed with the SLy5 force plus an added tensor component \cite{mar14}.
In this framework, investigation of the tensor effects on the s.p. energies and the first $J^{\pi}$=$2^{+}$ excited state can provide valuable information for the interpretation of the experimental results.

 In this work, as a first step we present the effect of the tensor force on the shell evolution of N=34 isotones:
$^{60}$Fe, $^{58}$Cr, $^{56}$Ti and $^{54}$Ca nuclei with both Hartree-Fock-Bogoliubov (HFB) and HF+BCS models using the 
Skyrme-type SLy5 (without tensor) \cite{cha98} and SLy5+T, T44\cite{les07} (with tensor) interactions. Among the many possible tensor parameter sets in the literature, we choose SLy5+T and T44 to illustrate the tensor force effects. The SLy5 interaction is designed for describing satisfactorily exotic nuclei \cite{cha98}. Furthermore, addition of a tensor force on top of SLy5 helps to explain the observed s.p. spectrum \cite{colo07}, which makes Sly5+T appropriate for illustrating the tensor force effects. To check the reliability of the results, we also use T44 which is another Skyrme interaction containing a tensor component.
Since the T44 interaction contains both the like-particle and proton-neutron tensor terms, it is also meaningful to study its predictions. The calculated s.p. energies with and without tensor force 
are compared at the mean field level. Then, the low-lying $J^{\pi}$=$2^{+}$ states are calculated within QRPA on top of HF+BCS for the Ca isotopic chain. Finally, conclusions are drawn on the possible magicity of $^{54}$Ca.

\section{Calculated single-particle spectra of N=34 isotones} 
In this section, we discuss the calculated shell evolution of the N=34 isotonic chain 
using the 
HFB method\cite{ben05}. In addition, 
we show that the HFB and HF+BCS methods yield very similar mean field properties, thus giving us more confidence in the quasi-particle (q.p.) RPA (QRPA) results presented in the next section.
The details of the HF+BCS and HFB calculations can be found in Refs. \cite{ben05,ring80}. We present 
the results obtained with and without the tensor force in order to investigate the magic nature of the $^{54}$Ca nucleus. 

 To study the effects of the tensor force on the s.p. energies of the N=34 isotonic chain, we compare the results obtained with SLy5, Sly5+T and T44 interactions. The SLy5+T and T44 interactions include tensor force components.
In our calculations, the tensor interaction parameters are 
$\alpha_{T}=$-170 MeV.fm$^{5}$, $\beta_{T}=$100 MeV.fm$^{5}$ for SLy5+T and 
$\alpha_{T}=$8.96 MeV.fm$^{5}$, $\beta_{T}=$113.02 MeV.fm$^{5}$ for T44. 
The parameters $\alpha_{T}$ and $\beta_{T}$ are given by:

\begin{equation}
\label{eq:pippo}
\alpha_{T}=\frac{5}{12}U
\quad\text{,}\quad
\beta_{T}=\frac{5}{24}(T+U)
\end{equation}
where T and U are the strengths of the original tensor force \cite{sky59}. In all calculations, we use a zero-range density-dependent pairing interaction \cite{ber91},

\begin{equation}
V_{pair}(\textbf{r}_{1},\textbf{r}_{2})=V_{0}\left[1-\eta\left(\frac{\rho(\textbf{r})}{\rho_{0}}\right)\right]\delta(\textbf{r}_{1}-\textbf{r}_{2})
\end{equation}
where $\rho(\textbf{r})$ is the particle density, $\rho_{0}=$0.16 fm$^{-3}$ is the nuclear saturation density and  we choose $\eta$$=$1 for the surface-type pairing interaction. The pairing strength ($V_{0}=$ 350 MeV. fm$^3$) is adjusted according to the experimental binding energies and two-neutron separation energies of the calcium isotopic chain. Additionally, the first 2$^{+}$ states of $^{42-46}$Ca isotopes have been used to obtain a proper pairing strength value. All Calculations are performed in a 20 fm box with 0.1 fm mesh size and a quasiparticle energy cut-off equal to 60 MeV. In our calculations, the maximum angular momentum value is taken as $j_{max.}=39/2$ ($j_{max.}=25/2$)
for neutrons (protons).
In the HFB approach the canonical quasiparticle energies $E_{i}$ are given by $E_{i}=\sqrt{(\varepsilon_{i}-\lambda)^{2}+\Delta_{i}^{2}}$ where $\varepsilon_{i}$ is the so-called Hartree-Fock equivalent single-particle energy, $\lambda$ is the chemical potential and $\Delta$ is the pairing gap which are defined in Ref. \cite{ben05}.

\begin{figure}[H]
  \begin{center}
    \includegraphics[width=4.6in]{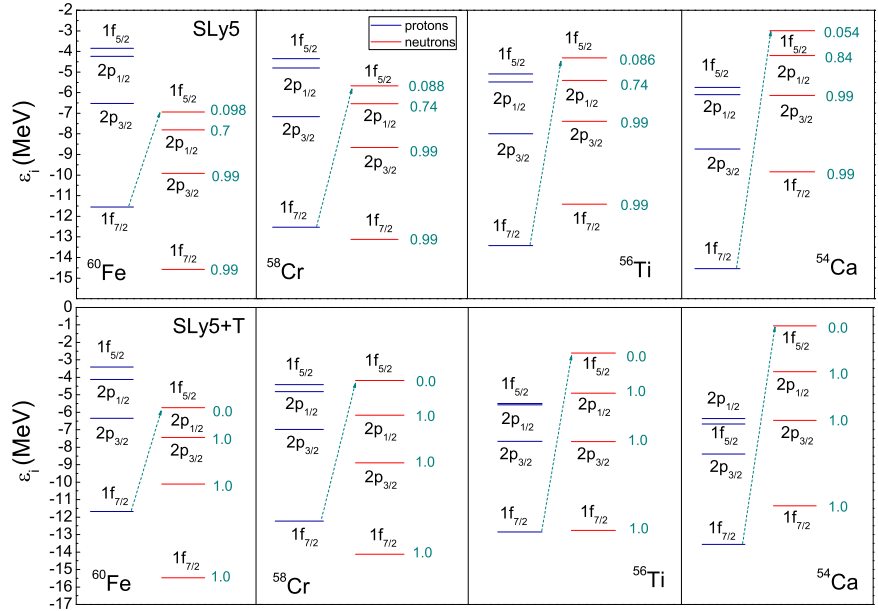}
  \end{center}
  \caption{Evolution of the $\varepsilon_{i}$ energies in N=34 isotones with two-proton removal in the nuclei $^{60}$Fe, $^{58}$Cr, $^{56}$Ti and $^{54}$Ca. The results with SLy5 and SLy5+T are displayed in the upper and lower panels, respectively. The occupation probabilities are shown for the neutron states.} 
  \label{100}
\end{figure}

 In the upper panel of Fig.\ref{100}, we show the $\varepsilon_{i}$'s in the N=34 isotones: $^{60}$Fe, $^{58}$Cr, $^{56}$Ti and $^{54}$Ca. They are calculated with the SLy5 interaction. With the removal of protons from $^{60}$Fe to $^{54}$Ca, the proton s.p. levels move down while the neutron s.p. levels move slightly up due to the increasing strength of the symmetry potential.
Also, the spin-orbit splittings of 
proton (resp. neutron) $p$ and $f$ states increase (resp. decrease). The energy gap between the last occupied ($2p_{1/2}$) and first unoccupied ($1f_{5/2}$) neutron states increases by about $0.4$ MeV when going from $^{60}$Fe to $^{54}$Ca. In $^{54}$Ca, the neutron gap energy is 1.2 MeV with SLy5 and the pairing effects are weak. 
The results indicate a non-magic $^{54}$Ca nucleus.

\begin{figure}[H]
  \begin{center}
    \includegraphics[width=4.6in]{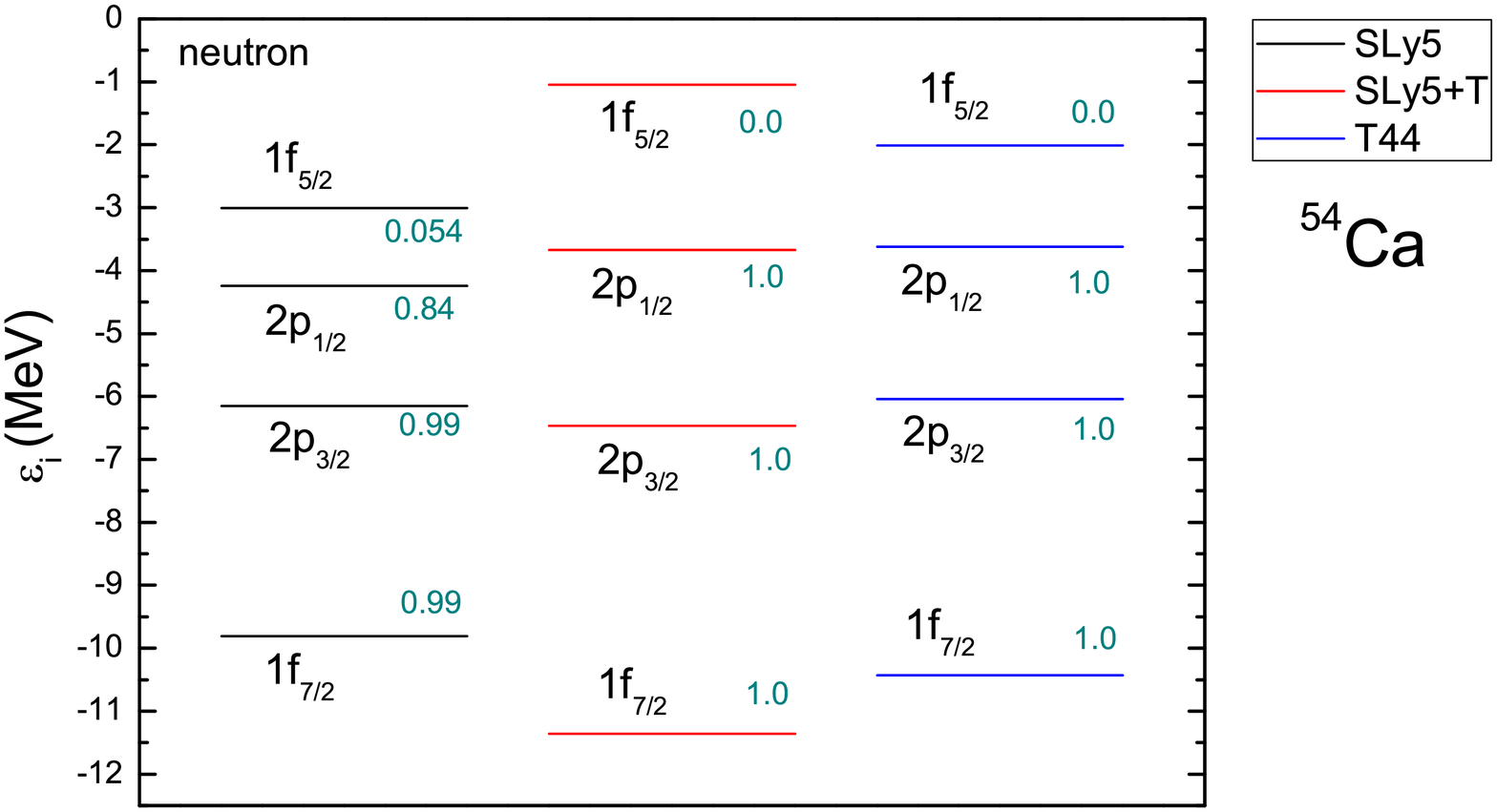}
  \end{center}
  \caption{Same as in Fig. \ref{100} for the neutron s.p. energies of $^{54}$Ca calculated with SLy5, SLy5+T and T44 interactions.} 
  \label{101}
\end{figure}

On the other hand, the results obtained with SLy5+T (lower panels of Fig.\ref{100}) are somewhat different.
They clearly indicate a neutron subshell closure at N=34. To better understand the influence of the tensor force on the neutron s.p. spectrum near the Fermi level in $^{54}$Ca, we show in Fig. \ref{101} the neutron s.p. levels for the $p$ and $f$ states. The HFB calculations are done with different Skyrme parametrizations, so as to check the model dependence of the results. Both SLy5+T and T44 interactions predict similar trends in the s.p. energies of N=34 isotones and they give no pairing in the $^{54}$Ca nucleus. Comparing SLy5 and SLy5+T spectra clearly shows that the tensor force enhances the occupied-unoccupied level gap. To check the model dependence of the results, we have performed the same calculations with the T44 interaction\cite{les07}. 
The results obtained with SLy5+T and T44 indicate that this feature is not specific to a given parametrization of the tensor force but it seems to be general. 

We conclude that the inclusion of the tensor component of the 
Skyrme interaction is necessary for explaining the experimental results about the magic nature of the $^{54}$Ca nucleus.
We have also compared our HFB results with those of the HF+BCS method, with and without tensor interaction. Both HFB and HF+BCS methods give practically the same predictions. The maximum relative difference between the s.p. energies of the two methods is about 1.0\%.

\section{Properties of the first $2^{+}$ state in Calcium isotopes}
Another clue about the influence of pairing effects on the shell structure of Ca isotopes is found in the evolution of the first $2^{+}$ state. Since the calculated q.p. spectra are very similar in HFB and HF+BCS descriptions, it is appropriate to perform QRPA calculations \cite{ring80} on top of the HF+BCS scheme in order to study this evolution.  
Our aim is to see the effects of the tensor force on the first 2$^{+}$ state and to compare with the data in Ca isotopic chain. 
In the QRPA part of the calculations, Skyrme energy density functionals are used in the derivation of the p-h matrix elements which also includes spin-orbit and two-body Coulomb interactions\cite{li12}. It should be noted that the tensor force is just used in the mean field part of the calculations but it is not included in the residual interaction part of the present QRPA study. However, the effect of the residual p-h tensor interaction is expected to be rather small in the $2^{+}$ excitations \cite{cao09,cao11}.

\begin{figure}[H]
    \includegraphics[width=5in]{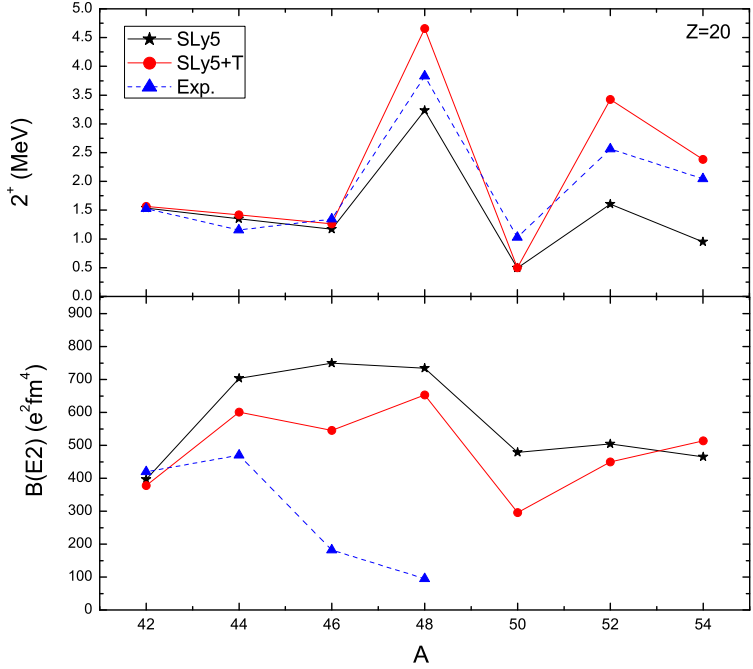}
  \caption{Upper panel: First $J^{\pi}$=$2^{+}$ energies in the  $^{42-54}$Ca isotopic chain with and without tensor force in the mean field part. Experimental values are taken from Ref.\cite{ra01} up to $^{52}$Ca and Ref. \cite{step13} for $^{54}$Ca nucleus. Lower Panel: B(E2) values. Experimental data are shown whenever available.}
  \label{102}
\end{figure}

The energies and B(E2) values of
the first $J^{\pi}$=$2^{+}$ states are shown in Fig. \ref{102}. 
In QRPA, the first $2^{+}$ states of the Ca isotopes 
are coming from single 2-q.p. excitations.
Inclusion of the tensor force has almost no effect from $^{42}$Ca to $^{46}$Ca and the predictions are in agreement with the data. However, the energies of the first $2^{+}$ states in $^{48}$Ca, $^{52}$Ca and $^{54}$Ca are increased by the tensor force. This increase is about 1.43 MeV, 1.82 MeV and 1.43 MeV, respectively. 
In addition, the first $2^{+}$ states of the $^{48}$Ca, $^{52}$Ca and $^{54}$Ca nuclei exhibit pure neutron s.p. excitations: ($1f_{7/2}$-$2p_{3/2}^{-1}$), ($2p_{3/2}$-$2p_{1/2}^{-1}$) and ($2p_{1/2}$-$1f_{5/2}^{-1}$), respectively.

The first $2^{+}$ state of $^{54}$Ca nucleus has been discussed before in shell model approach \cite{rej07,hol13} which are not able to predict the observed energies in Ref. \cite{step13}. In $^{54}$Ca, we obtain a good agreement with the latest experimental results. Inclusion of the tensor force in the mean field part of the calculations explains well the behavior of the first $2^{+}$ states in this nucleus.

 In the lower panel of Fig. \ref{102}, we show the B(E2) values calculated with and without the tensor interaction. The tensor force does not modify much the B(E2) values, as expected. However,  the QRPA B(E2) values with and without tensor force are not in agreement with the data. This is a general drawback of RPA and QRPA that the electromagnetic transition probabilities of collective low-lying states are generally not well described, even in stable nuclei. This may be related to the property that the RPA approach is better suited for describing collective modes corresponding to an irrotational flow in a hydrodynamic picture \cite{saga83}, which is the case for giant resonances but not so much so for low-lying electric modes.  

\section{Summary}
In this work, we have employed both HFB and HF+BCS methods to study the ground states of N=34 isotones with the aim of exploring the possible magic nature of $^{54}$Ca.  
The calculations were performed with Skyrme type interactions, without (SLy5) and with a tensor component (SLy5+T, T44).
This choice is guided by the fact that the tensor force affects the s.p. spectra and thus, it can change the magicity behavior of those nuclei.
The evolution of the s.p. spectra was explored. 
We do not obtain a clear sign of sub-shell closure with the SLy5 interaction, but the addition of a tensor force component quite enhances the energy difference between the s.p. states, 
thus leading to
a subshell closure in $^{54}$Ca. 
We have also investigated the first 
$2^{+}$ states in the Ca isotopic chain, again with and without the tensor component in the Skyrme interaction.
The tensor force leads to an increased energy of the first $2^{+}$ state in the calcium isotopic chain. Both $^{52}$Ca and $^{54}$Ca show a magic nature. Especially, the first $2^{+}$ state of $^{54}$Ca is well reproduced
when compared with the experimental results. However, the B(E2) values are not well reproduced, 
as it is often the case with QRPA applied to low-lying collective states in spherical nuclei. 
The results indicate the importance of the tensor force on the shell evolution of $p-f$ shell nuclei.

\end{document}